\documentclass[prl,twocolumn,showpacs,nofootinbib,superscriptaddress]{revtex4}

\usepackage{bm,dcolumn,amsmath,graphicx,amsfonts,amssymb}
\usepackage{epsfig}

\begin{document}
\title{ Highly-charged ions as a basis of  optical atomic clockwork of exceptional accuracy}

\author{Andrei Derevianko}
\affiliation{Department of Physics, University of Nevada, Reno,
Nevada 89557, USA}

\author{ V.A. Dzuba}
\affiliation{Department of Physics, University of Nevada, Reno,
Nevada 89557, USA}
\affiliation{
School of Physics, University of New South Wales, Sydney, NSW 2052, Australia}

\author{V. V. Flambaum}
\affiliation{
School of Physics, University of New South Wales,
Sydney, NSW 2052, Australia}

\begin{abstract}
We propose a novel class of atomic clocks based on highly charged ions.
We consider highly-forbidden laser-accessible  transitions within the $4f^{12}$ ground-state
configurations of  highly charged ions. Our evaluation of systematic effects demonstrates  that these
transitions may be used for building exceptionally accurate atomic clocks which may
compete in accuracy with recently proposed nuclear clock. 
\end{abstract}

\date{ \today }

\pacs{06.30.Ft,32.10.-f}
\maketitle

Atomic clocks are arguably the most precise scientific instruments ever built.
Their exquisite  precision  has enabled both foundational tests  of modern physics, e.g., probing hypothetical drift of fundamental constants~\cite{RosHumSch08}, and practical applications, such as the global positioning system. 
State of the art clocks carry out frequency measurements at the eighteenth decimal place~\cite{ChoHumKoe10}.
As  the projected  fractional accuracy of such clocks is at the level of $10^{-18}$~\cite{DerKat11,RosSchHum07}, it is natural to wonder how to extend the  accuracy frontier even further. 
We are only aware of one proposal, the nuclear clock~\cite{CamRadKuz12},  that holds the promise of  reaching  the improved $10^{-19}$  accuracy level. 
The nuclear clock, however, relies on a yet unobserved   optical transition in the radioactive $^{229}\mathrm{Th}$ nucleus. 
Here we show that the nuclear clock performances can be replicated with atomic systems, fully overcoming  these challenges.  We identify several highly-forbidden laser-accessible  transitions in heavy stable isotopes of highly-charged ions (HCI) that may serve as clock transitions.  Similarly to  the  singly-charged ions of modern clocks~\cite{ChoHumKoe10}, HCIs  can be trapped and cooled~\cite{GruHolSch05,HobSolSuh11}. The key advantage of HCIs comes from their higher ionic charge.
As the ionic charge  increases,  the electronic cloud shrinks thereby  greatly reducing couplings to detrimental external perturbations.  Our analysis of various systematic effects for several HCIs demonstrates the feasibility of attaining the $10^{-19}$ accuracy mark with existing technology.

Atomic clocks operate by locking the frequency of  external oscillator (e.g., laser cavity) to a quantum (atomic/nuclear/ionic/molecular) transition. One tells time by simply  counting the number of oscillations at the source and multiplying it by the known oscillation period. A suitable clock transition should have a good quality factor (Q-factor). Moreover,  the clock transition frequency must remain unaffected by external perturbations. Finally, one has to be able to interrogate quantum oscillators for a long time,  so the atoms should be trapped.  The clock stability and accuracy generally improve with higher frequency of the clock transition, $\nu_\mathrm{clock}$, and the current accuracy record~\cite{ChoHumKoe10} is  held by singly-charged ion clocks operating at optical frequencies. 


Before we start with the clock estimates, we would like to recapitulate a few basic facts 
about HCIs. 
In a multi-electron atom,  optical electrons move in mean-field potential created by other electrons and the nucleus. 
However, as the electrons are stripped away from the atom, the field experienced by the optical electrons becomes increasingly  coulombic, and 
one could invoke intuitive hydrogen-ion-like estimates~\cite{Gil01}. For example,  the size of the electron cloud scales  with the residual nuclear (ionic) charge $Z_i$ as $1/ Z_i$. Since typical matrix elements are proportional to some power of atomic radius, most of the couplings to the detrimental external perturbations scale down with increasing $Z_i$.  Higher-order responses, e.\ g., polarizabilities, are suppressed even further due to increasing energy intervals that scale as $Z_i^2$.   Such suppression of couplings to external perturbations is the key to improved accuracy in the proposed HCI clock.

Trapping and cooling clock ions beneficially  increases interrogation time and reduces Doppler shifts.
HCIs can be loaded in ion traps~\cite{GruHolSch05,HobSolSuh11}, however, due to the $Z_i^2$ scaling of intervals most of HCIs lack low-energy electric dipole transitions that can be used
for direct laser cooling. As in the state-of-the-art optical ion clocks~\cite{ChoHumKoe10}, to circumvent this limitation, one may choose to employ
sympathetic cooling.  In this scheme, long-range elastic Coulomb collisions with
continually laser-cooled Be$^{+}$ ions drive HCI temperature down to
mK temperatures. 
Heavy HCIs may be cotrapped
with relatively light ions of low ionic charge, because equations of motion in ion traps depend only on the ratio of ion charge to its mass, $Z_{i}/M$. 
For example, Ref.~\cite{GruHolSch05} experimentally demonstrated sympathetic cooling of Xe$^{44+}$ with Be$^+$ ions.  
The basic idea~\cite{ChuSchSte99} is to initially cool HCIs resistively and then load
precooled HCIs into the Be$^{+}$ ion trap. At sufficiently low temperatures the
rates of undesirable  charge-exchange processes between two ionic
species become negligible. 
Heavier cooling species like
Mg$^+$ can be also used~\cite{ChoHumKoe10} to improve mass-matching and thereby the cooling efficiency. 
Additional advantage of co-trapping two ionic species comes from the
possibility of carrying out quantum logic spectroscopic clock readout~\cite{ChoHumKoe10}
and initialization.

\begin{figure}
\centering
\epsfig{figure=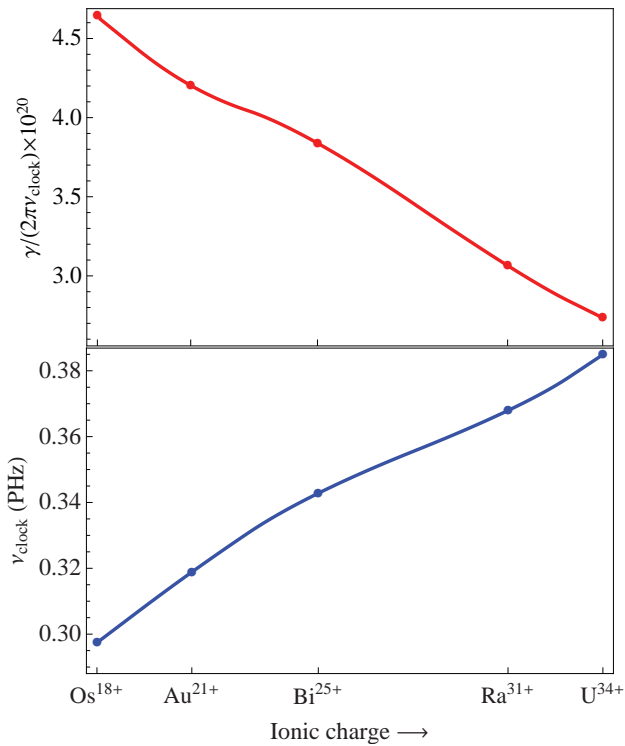,scale=0.65}
\caption{ Systematic trend of basic clock properties for the  [Pd]$4f^{12}$ isoelectronic sequence. Clock frequencies (lower panel) and the ratios of radiative line width to the clock frequency  (upper panel) are shown  as a function of ionic charge. Clock frequencies lie in laser-accessible near-infrared and optical domains,
while the narrow transition width assures high quality factor. }
\label{Fig:Overview}
\end{figure}

There are many possible choices of ions. Including all degrees of ionization of  the first 112 elements of the Mendeleev  (periodic)  table leads
to 6,216 potential ions.  
We are interested in those ions which have closed  highly-forbidden transitions in the optical frequency band. Some of the optical transitions in HCIs were identified in Refs.~\cite{BerDzuFla10,BerDzuFla11,BerDzuFla12}. We  analyzed several possibilities, and we find HCIs with the  [Pd]$4f^{12}$ ground-state electronic configuration to be especially promising for precision timekeeping. 

The [Pd]$4f^{12}$ configuration is the ground state configuration for all ions starting from
Re$^{17+}$ which have nuclear charge $Z \geq 75$ and ionic charge $Z_i =
Z-58$.   We computed properties of such ions using relativistic configuration interaction method described in
\cite{DzuFla08a,DzuFla08b};  details of the calculations will be presented elsewhere.
According to the Hund's rules, in these HCIs  the $4f^{12} \,^3\!\mathrm{H}_6$ and $4f^{12}\, ^3\!\mathrm{F}_4$  states are the ground  and the first excited states respectively.  Our computed clock frequencies $\nu_\mathrm{clock}$ and ratios of radiative width $\gamma$ to  $\nu_\mathrm{clock}$  are shown  in Fig. ~\ref{Fig:Overview}.  The transition frequencies range from the near-infrared to the optical region and are laser-accessible.  The clock states are exceptionally narrow, as the upper clock state may decay only via highly-suppressed   electric-quadrupole (E2)
transition. The resulting lifetime of a few hours  leads to the prerequisite  high Q-factor of the clock transition, so that $\gamma/( 2\pi \nu_\mathrm{clock})$ remains below the $10^{-19}$ accuracy goal.


Clock-related properties of several representative HCIs are compiled in Table~\ref{Tab:Properties}.
As an example, below we focus on $^{209}$Bi$^{25+} $ ion. It has the highest transition frequency and  Q-factor among stable isotopes. Being the heaviest among such isotopes has additional advantages as its large mass suppresses systematic effects related to Doppler shifts (see below).

\begin{table}
\caption{Clock-related properties of representative HCIs of the  [Pd]$4f^{12}$ isoelectronic sequence.
$\lambda_{\mathrm{clock}}$ is the wavelength of the clock transition, $\tau$ is the lifetime of the upper $4f^{12}\, ^3\!\mathrm{F}_4$ clock level, and 
$Q$ is the quality factor. Systematic clock shifts are governed by differential static electric-dipole polarizability $\Delta\alpha^{E1}\left(  0\right)$, black-body coefficient $\beta_{\mathrm{BBR}}$ and quadrupole moments of the clock states $Q^e$. Hyperfine structure of the clock levels is determined by the nuclear spin $I$ and hyperfine structure constants $A$.
Numbers in square brackets represent powers of 10.} 
\label{Tab:Properties}
\begin{ruledtabular}
\begin{tabular}
[c]{llll}
&  \multicolumn{1}{c}{$^{189}$Os$^{18+}$} & 
 \multicolumn{1}{c}{$^{209}$Bi$^{25+}$} & 
  \multicolumn{1}{c}{$^{235}$U$^{34+}$}\\
\hline
$\lambda_{\mathrm{clock}}$, nm & 1010 & 874 & 779\\
$\tau$, hrs & 3.2 & 3.4 & 4.2\\
$1/Q$ & 4.6[-20] & 3.8[-20] & 2.7[-20]\\[1ex]
$\Delta\alpha^{E1}\left(  0\right)  ,a_{0}^{3}$ &  -2.3[-3] & -2.3[-4] &
-8[-5]\\
$\beta_{\mathrm{BBR}}$ & 6.6[-20] & \ 5.8[-21] & 1.8[-21]\\
$Q^{e}\left(  ^{3}H_{6}\right)  ,\left\vert e\right\vert a_{0}^{2}$ & 1.84[-1] &
1.24[-1] & 8.3[-2]\\
$Q^{e}\left(  ^{3}F_{4}\right)  ,\left\vert e\right\vert a_{0}^{2}$ &
-1.51[-2] & -1.24[-2] & -8.4[-3]\\[1ex]
$I$ & 3/2 & 9/2 & 7/2\\
$A\left(  ^{3}H_{6}\right)  $, MHz & 688 & 2523 & -484\\
$A\left(  ^{3}F_{4}\right)  $, MHz & 719 & 2584 & -493
\end{tabular}
\end{ruledtabular}
\end{table}

%


\begin{figure}
\centering
\epsfig{figure=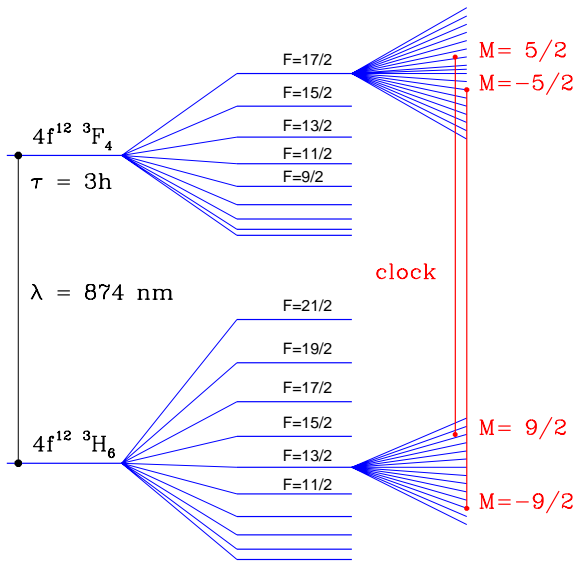,scale=0.45}
\caption{Proposed virtual clock transitions in  $^{209}$Bi$^{25+}$ highly-charged ion. Averaging over the two indicated  transition frequencies removes the first-order Zeeman shift. Specific choice of hyperfine components and magnetic sub-states minimizes shifts due to couplings to  gradients of the trapping electric field.   
}
\label{Fig:ClockTransionBi}
\end{figure}

$^{209}$Bi has the nuclear spin of $I=9/2$ and, as sketched in Fig.~\ref{Fig:ClockTransionBi},   the electronic  states are split into a multitude of hyperfine components.
The clock transitions must be insensitive to external perturbations, such as magnetic and electric fields. Because of that we choose specific hyperfine states and magnetic substates: $|F=17/2, M_F=\pm 5/2\rangle \leftrightarrow |F=13/2, M_F=\pm 9/2\rangle$ for clock transition.
 Similar to the virtual transition technique demonstrated in Hg$^+$ clocks~\cite{RosHumSch08}, 
the HCI clock operates on two transitions that  have opposite g-factors. Averaging over the two transition frequencies eliminates the linear Zeeman shift, making clock insensitive to B-fields.
While such a technique could be applied to multiple transitions, as shown below, we further required that  our specific choice  minimizes couplings to electric field gradients.

 Clock accuracy is affected by multiple systematic effects: magnetic fields, electric fields, Doppler (motion-induced) effects, blackbody radiation (BBR), and gravity. We show that all these effects are suppressed at the desired $10^{-19}$ fractional accuracy. 

We start with examining the BBR shifts; these arise due to perturbations by the  photon bath at room temperature. The fractional shift reads~\cite{PorDer06}
\begin{equation*}
\frac{\Delta\nu_{\mathrm{BBR}}}{\nu_{\mathrm{clock}}}   \approx-\frac{\pi
^{2}}{15c^{3}\hbar^{4}}\frac{\left(  k_{B}T\right)  ^{4}}{\nu_{\mathrm{clock}%
}}~\Delta\alpha(0) \equiv \beta_\mathrm{BBR} \times \left( \frac{T}{300\, \mathrm{K}} \right)^4 \, ,
\end{equation*}
where $\Delta\alpha(0)$ is the differential static polarizability of the clock transition and $T$ is the BBR temperature.  For HCIs, polarizabilities are suppressed as $1/Z_i^4$ and our calculation yields $\Delta\alpha(0)\approx -2.3 \times 10^{-4}\, a_0^3$ ($a_0$ is the Bohr radius). 
This tiny $\Delta\alpha(0)$ translates into
 the fractional BBR shift at room temperature of just $5.8 \times 10^{-21}$. 
Similarly, differential polarizability determines sensitivity to stray electric fields: $\Delta \nu /\nu _{\mathrm{clock}}=-\,\Delta \alpha (0)\mathcal{E}^{2}/\left(
2h\nu _{\mathrm{clock}}\right)$.
Typical E-fields of $10 \, \mathrm{V/m}$ lead to negligible $10^{-284}$-level shifts. Cooling lasers shining on the coolant ion will lead to AC Stark shifts of the HCI clock levels. Again compared to the singly-charged ion clocks these shifts will be strongly suppressed due to much smaller $\Delta\alpha$ and also because the  HCI and the coolant ion are  repelled by stronger Coulomb forces reducing the overlap of the cooling laser beam with the HCI. 

The clock ion is trapped in a non-uniform field; the gradient of this field couples to the quadrupole moment $Q$ of the clock states~\cite{Ita00}. 
The quadrupole shift (QS) of the
clock transition is given by
\begin{equation}
\frac{\Delta\nu_{QS}}{\nu_\mathrm{clock}}=-\frac{1}{2h\nu_\mathrm{clock}}\Delta Q  \frac{\partial \mathcal{E}_{z}}{\partial z} \, ,  \label{Eq:QS}
\end{equation}
where $\Delta Q \equiv \langle Q_{0}\rangle_{e}-\langle
Q_{0}\rangle_{g}$ is the difference in expectation values of the zeroth
component of the quadrupolar tensor for the upper and lower clock states.

\begin{figure}
\centering
\epsfig{figure=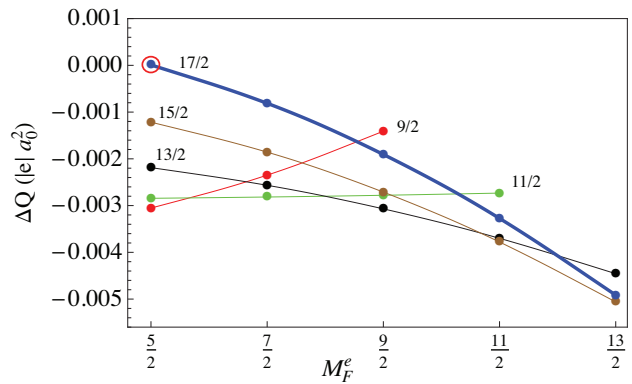,scale=0.65}
\caption{ Differential quadrupole moment $\Delta Q$ as a function of magnetic quantum number of the excited clock level for $^{209}$Bi$^{25+}$.  
Different curves are labeled by values of the total angular momentum of the excited state $F$. The ground state hyperfine component remains fixed $|F_g=13/2, M_g =  9/2 \rangle$. The minimial value of $\Delta Q$ is  encircled; it is attained for the $|F_e=17/2, M_e= \pm 5/2 \rangle$ 
``magic''  state.
}
\label{Fig:dQSearch}
\end{figure}

The relevant clock shifts for singly-charged ion clocks are sizable and considerable efforts has been devoted to mitigating this effect~\cite{Ita00,RosHumSch08}.
While for the HCIs one expects $1/Z_i^2$ suppression of $Q$-moments, we find that the relevant clock shifts can be still appreciable. Below we minimize 
the quadrupole shift  by exploiting the richness of the hyperfine structure of the [Pd]$4f^{12}$ HCI  clock states.
Indeed, the Q-moments of various hyperfine substates $|\alpha JI;FM_{F}%
\rangle$ attached to electronic state $|\alpha J\rangle$ may be conveniently
expressed as a product of  Q-moment of the electronic state
$Q^{e}\left(  \alpha J\right)  $ and a kinematic factor
\begin{eqnarray*}
\lefteqn{\langle\alpha JI;FM_{F}|Q_{0}|\alpha JI;FM_{F}\rangle= } \\
&& \left(  3M_{F}%
^{2}-F\left(  F+1\right)  \right)  \, K\left(  J,I,F\right)  \, Q^{e}\left(
\alpha J\right)  \,
\end{eqnarray*}
where $Q^e$ are listed in Table~\ref{Tab:Properties} and the  $M_{F}$-independent factor $K\left(  J,I,F\right)$   reads
\begin{eqnarray*}
\lefteqn{ K\left(  J,I,F\right)  =(-1)^{J+I+F}\frac{2F+1}{2J\left(  2J-1\right)} \times } \\
&&\left\{
\begin{array}
[c]{ccc}%
J & F & I\\
F & J & 2
\end{array}
\right\}  \left(  \frac{\left(  2J+3\right)  !}{\left(  2F+3\right)  !}%
\frac{\left(  2F-2\right)  !}{\left(  2J-2\right)  !}\right)  ^{1/2}.
\end{eqnarray*}

As  the gradient is fixed by trap parameters,  we minimize the difference   $ \langle Q_{0}\rangle_{e}-\langle
Q_{0}\rangle_{g}$ by considering all possible pairs of  hyperfine sub-states allowed by the E2 selection rules. 
The search for ``magic'' transitions
depends on the ratio of electronic Q-moments. These ratios can be determined experimentally by measuring frequencies of several hyperfine transitions in a trapped ion. 
In Fig.~\ref{Fig:dQSearch} we illustrate such a search based on our computed values of Q-moments for $^{209}$Bi$^{25+}$. 
We find that  the minimal value of $\Delta Q = -5 \times 10^{-6} |e|a_0^2 $ is attained for the magnetic components $|F_e=17/2, M_e= \pm 5/2 \rangle - |F_g=13/2, M_g = \pm 9/2 \rangle$.  These are the clock transitions indicated   in Fig.~\ref{Fig:ClockTransionBi}. 

The field gradient  in Eq.~(\ref{Eq:QS}) is fixed by the trap; we adopt the value from  the Be$^+$/Al$^+$ clock~\cite{TillPrivateComm}:  $\partial \mathcal{E}_{z} / \partial z \approx 10^8 \, \mathrm{V}/\mathrm{m}^2$.  Additional gradient on the clock HCI is exerted  by the coolant ion. For typical ion separations of 10 $\mu$m the resulting gradients are smaller than the indicated trapping field gradient. With such gradients, 
we find  ${\Delta\nu_{QS}}/{\nu_\mathrm{clock}} \approx 5 \times 10^{-19}$, which can be substantially reduced further. Indeed,  due to rotational symmetry arguments, the QS can be fully zeroed out by averaging  clock measurement  over three orthogonal directions of quantizing B-field~\cite{Ita00}. 
The power of this technique has been experimentally demonstrated for the Hg$^+$ 
clock~\cite{RosHumSch08}, where the QS was reduced by a factor of 200. The averaging out was not exact due to technical alignment issues. Combination of the ``magic'' choice of clock states with the averaging technique leads to our projected QS uncertainty of $\Delta \nu_{QS}/ \nu_\mathrm{clock} = 
 2.5 \times 10^{-21}$.

Clock frequencies are affected by magnetic fields. The first-order Zeeman
shift can be eliminated by averaging the measurements over two virtual clock transitions indicated in Fig.~\ref{Fig:ClockTransionBi}. The dominant source of Zeeman-related uncertainties comes from AC B-fields caused by
currents at the RF trap frequencies in conductors near the ion. Ideally, such fields would vanish at the trap axis, but in practice  $B_\mathrm{AC}$ is always present due to geometric imperfections~\cite{RosHumSch08}. We adopt  $B_\mathrm{AC}=5\times 10^{-8}\, \mathrm{T}$ measured in the 
Al$^+$/Be$^+$ trap~\cite{RosHumSch08} as the typical value. The AC fields contribute to the second-order Zeeman shift. Calculations of the relevant differential magnetic-dipole polarizability $\Delta \alpha^\mathrm{M1}$ are dominated by intermediate states of clock hyperfine manifolds. For our choice of magnetic substates for $^{209}$Bi$^{25+}$ , we find  $\Delta \alpha^\mathrm{M1} \approx  -2.1 \times 10^{10} \, \mathrm{Hz}/\mathrm{T}^2$ which translates into a fractional clock shift of $4 \times 10^{-20}$ ; it  is below the sought accuracy goal.

Working with HCI requires relatively high vacuum attainable in cryogenic traps cooled with liquid helium.
In the context of ion clocks, such traps were demonstrated for Hg$^+$ ion~\cite{RosHumSch08}. The rate coefficient~\cite{FerKorVer97}  for charge-exchange collisions of heavy HCIs with residual  He atoms scales as $Z_i$: $k\approx 0.5 \times 10^{-9}  Z_i  \, \mathrm{cm^3}/\mathrm{s}$.  If the HCI were to survive for an hour, the number density of He atoms would have to be limited by $2 \times 10^4 \, \mathrm{cm}^{-3}$.

The zero-point-energy  motion of trapped ion has a profound effect on the  clock
accuracy via the effect of special relativity, time dilation~\cite{ChoHumRos10}. The fractional
effect of time dilation may be evaluated as the ratio of the ion kinetic
energy $K$ to its rest mass energy,
\begin{equation}
\delta\nu_{\mathrm{TD}}/\nu_{\mathrm{clock}}=-K/Mc^{2}. 
\label{Eq:TD}
\end{equation}

To estimate the effects of time-dilation, we adopt trap parameters from the
ion clock of Ref.~\cite{RosSchHum07} based on a pair of Al$^{+}$/Be$^{+}$
ions. Indeed, once the trapping fields are specified, ion motion is entirely
characterized by the ratio of ionic charge to its mass, $Z_{i}/A$, where $A$
is the atomic weight. For all the enumerated highly-charged clock ions, this
ratio is about 0.15 which is comparable to the $Z_{i}/A$-ratio for Be$^{+}$.
Since the trapping parameters are similar, the value of kinetic energy remains
roughly the same as in the Al$^{+}$/Be$^{+}$ clock, while the enumerated HCIs
are about 10 times heavier than Al. This mass difference leads to suppression
of the time-dilation effects with heavy HCIs, see Eq.(\ref{Eq:TD}). In the demonstrated $^{27}$Al$^{+}$
clocks the uncertainties due to time-dilation are at the level of a few parts
in $10^{-18}$ with the  goal of reaching the $10^{-18}$ accuracy
milestone. Due to the mass scaling argument we anticipate that $10^{-19}$ is
the plausible accuracy goal for the proposed HCI clocks. Notice that this limitation is not fundamental as it relates to the technical ability to control stray electric fields in the trap.

The clocks are affected by the effects of general relativity as well~\cite{ChoHumRos10}. 
The fractional frequency difference between two clocks at differing heights on Earth's surface is $\Delta \nu_G/\nu_\mathrm{clock}$\,=\,$g \, \Delta h/c^2$, where $g$ is the gravitational acceleration  and $\Delta h$ is the  difference in clock height positioning.  If the two identical clocks differ in height by 1 mm, the clock would acquire  a 1$\times$$10^{-19}$ fractional frequency shift. Such uncertainty would limit accuracy of time transfer.

To summarize, we have shown that the highly-charged ions  may serve as a basis of optical atomic clockwork at the $10^{-19}$ fractional accuracy. 
 Such accuracy results from 
the smallness of  electronic cloud in ions and therefore suppressed  couplings to external perturbations. The  $10^{-19}$ fractional accuracy is  matched only by the proposed nuclear clock~\cite{CamRadKuz12}; our proposed clock avoids complications of radioactivity and  uncertainties in transition frequencies
associated with the nuclear clock.



We would like to thank P. Beiersdorfer, T. Rosenband, E. Peik, J. Weinstein, and D. Wineland for discussions.
The work was supported in part by the U.S.  National Science Foundation and by the  Australian Research Council.


\end{document}